\documentclass{article}
\usepackage{amssymb}
\usepackage{amscd} 
\newtheorem{theorem}{Theorem}
\newtheorem{proposition}[theorem]{Proposition}
\newtheorem{lemma}[theorem]{Lemma}
\newtheorem{remark}[theorem]{Remark}
\begin{document} 
\begin{center}
\textbf{\Large Soliton Solutions for the Non-autonomous Discrete-time Toda 
 Lattice Equation}\\[10mm]
{\large Kenji Kajiwara$^{1,2}$ and Atsushi Mukaihira$^1$}\\[5mm]
{\large $^1$: Graduate School of Mathematics, Kyushu University,\\[2mm]
6-10-1 Hakozaki, Fukuoka 812-8581, Japan} \\[3mm]
{\large $^2$: School of Mathematics and Statistics F07, University of Sydney,\\[2mm]
Sydney, NSW 2006, Australia} 
\end{center}
\begin{abstract} 
We construct $N$-soliton solution for the non-autonomous discrete-time Toda lattice
 equation, which is a generalization of the discrete-time Toda equation
 such that the lattice interval with respect to time is an arbitrary
 function in time.
\end{abstract}
\section{Introduction}
In this article, we consider the nonlinear partial difference equation given by
\begin{equation}
\begin{array}{l}
\medskip
{\displaystyle A_n^{t+1}+B_n^{t+1}+\lambda_{t+1}=A_n^t+B_{n+1}^t+\lambda_t,} \\
{\displaystyle A_{n-1}^{t+1}B_n^{t+1}=A_n^tB_n^t, }
\end{array}\quad n\in\mathbb{Z},\label{dTL}
\end{equation}
where $n$ and $t$ are the independent variables, $A_n^t$ and $B_n^t$ are
the dependent variables and $\lambda_t$ is an arbitrary function in $t$,
respectively.  In the physical context, the variables $n$, $t$, $A_n^t$
and $B_n^t$ are identical as the lattice site, the discrete time and the
fields, respectively.  Equation (\ref{dTL}) is equivalent to the following equation
\begin{equation}
\begin{array}{l}
\medskip
{\displaystyle  J_n^{t+1}-\delta_{t+1}V_{n-1}^{t+1}=J_n^t-\delta_t V_n^t,}\\
{\displaystyle  V_n^{t+1}(1-\delta_{t+1} J_n^{t+1}) =V_n^t(1-\delta_t J_{n+1}^t),}
\end{array}\quad n\in\mathbb{Z},\label{RTL}
\end{equation}
where the variables are related as
\begin{equation}
 A_n^t=-\lambda_t+J_n^t,\quad B_n^t=-\lambda_t^{-1}V_{n-1}^t,\quad \delta_t=\lambda_t^{-1},
\end{equation}
respectively. When $\lambda_t$ or $\delta_t$ is a constant, equation
(\ref{dTL}) or (\ref{RTL}) reduces to the discrete-time Toda equation
proposed by Hirota \cite{Hir1977, HIK1988}. Moreover, equation
(\ref{RTL}) yields the celebrated Toda lattice equation 
\begin{equation}
\begin{array}{l}
\medskip
{\displaystyle  \frac{dJ_n}{dt}=V_{n-1}-V_n,}\\
{\displaystyle  \frac{dV_n}{dt}=V_n(J_n-J_{n+1}),}
\end{array}\quad n\in\mathbb{Z},\label{TL:0}
\end{equation}
in the continous limit $\delta_t=\delta\to 0$.

Equation (\ref{dTL}) was proposed by Spiridonov and Zhedanov in
\cite{SZ1995}, where the equation is called as just ``the discrete-time
Toda lattice''. On the other hand, equation (\ref{RTL}) was proposed by
Hirota\cite{Hir1997}, and called as ``the random-time Toda equation''.
However, it appears that those names are not appropriate for equations
(\ref{dTL}) and (\ref{RTL}), since the former name usually refers the
case where $\lambda_t$ and $\delta_t$ are constants, and the latter is
somewhat misleading.  In this article, we call equations (\ref{dTL}) and
(\ref{RTL}) as ``the non-autonomous discrete-time Toda lattice equation''.
%
%

The non-autonomous discrete-time Toda lattice equation is written in the Lax form
\begin{equation}
 L_{t+1}R_{t+1}+\lambda_{t+1}=R_t L_t+\lambda_t,
 \label{Lax eq}
\end{equation}
where $L_t$ and $R_t$ are difference operators defined by
\begin{equation}
 L_t=A_n^t+e^{-\partial_n},\quad R_t=B_{n+1}^t e^{\partial_n}+1,
\end{equation}
respectively. The Lax equation (\ref{Lax eq}) is the compatibility condition of the
spectral problem equation
\begin{equation}
\begin{array}{l}
\medskip
{\displaystyle  \Psi_n^{t+1}=R_t \Psi_n^t=B_{n+1}^t \Psi_{n+1}^t+\Psi_n^t,}\\
{\displaystyle  (x-\lambda_t)\Psi_n^t=L_t \Psi_n^{t+1}=A_n^t\Psi_n^{t+1}+\Psi_{n-1}^{t+1},}
\end{array}
\label{Spec eq 1}
\end{equation}
where $x$ is a spectral parameter and $\Psi_n^t$ a wave function.

An important feature of soliton equations, including
the Toda lattice and the discrete-time
Toda equations is that they admit wide class of exact solutions, such as
soliton solutions.
Moreover, these solutions are expressed by determinants or
Pfaffians\cite{Hir2004}, which are regarded as
characteristic property of integrable systems according to the Sato theory\cite{MJD1999}.
It is known that the discrete-time Toda equation (when $\lambda_t$ is a
constant) admits two kinds of determinant solutions.
One is the Hankel type determinant solution, in which the lattice site
$n$ appears as the determinant size\cite{Hir1987,KMNOY2001}.
Another one is the Casorati determinant solution which describes
soliton type solutions\cite{HIK1988}.
In this solution, the determinant size corresponds to the number of
solitons.
The Hankel type determinant solution for the non-autonomous discrete-time
Toda lattice equation on the semi-infinite lattice was constructed in
\cite{MT2004a,MT2004b}.
The purpose of this article is to present explicit $N$-soliton solutions
for the non-autonomous discrete-time Toda lattice equation in the form of
Casorati determinant.
\section{Soliton solution for the non-autonomous discrete-time Toda lattice equation}\label{Casorati}
For any $N\in\mathbb{Z}_{>0}$, we first define $N\times N$ Casorati
determinants $\tau_n^t$ and $\sigma_n^t$ as
\begin{eqnarray}
\label{sol tau fns}
 \tau_n^t&=&
 \left|\begin{array}{cccc}
  \varphi_1^t(n) & \varphi_1^t(n+1) & \cdots & \varphi_1^t(n+N-1)\\
  \varphi_2^t(n) & \varphi_2^t(n+1) & \cdots & \varphi_2^t(n+N-1)\\
  \vdots & \vdots & & \vdots\\
  \varphi_N^t(n) & \varphi_N^t(n+1) & \cdots & \varphi_N^t(n+N-1)
\end{array}\right|,\\
 \sigma_n^t&=&
 \left|\begin{array}{cccc}
  \psi_1^t(n) & \psi_1^t(n+1) & \cdots & \psi_1^t(n+N-1)\\
  \psi_2^t(n) & \psi_2^t(n+1) & \cdots & \psi_2^t(n+N-1)\\
  \vdots & \vdots & & \vdots\\
  \psi_N^t(n) & \psi_N^t(n+1) & \cdots & \psi_N^t(n+N-1)
\end{array}\right|,
\end{eqnarray}
where the entries $\varphi_i^t(n)$ and $\psi_i^t(n)$ ($i=1,\ldots,N$) satisfy linear
relations
\begin{eqnarray}
&& \varphi_i^{t+1}(n)=\varphi_i^t(n)-\mu_t \varphi_i^t(n+1),
 \label{sol disp rel 1}\\
&& \psi_i^t(n)=\varphi_i^{t-1}(n)-\mu_t \varphi_i^{t-1}(n+1),
 \label{sol disp rel 2}\\
&& P_i^t \varphi_i^{t-1}(n)=\psi_i^t(n)-\mu_t \psi_i^t(n-1),
 \label{sol disp rel 3}
\end{eqnarray}
with $\mu_t$ being an arbitrary function of $t$, and $P_i^t$
given by
\begin{equation}
 P_i^t=(1-p_i \mu_t)(1-p_i^{-1}\mu_t),\quad i=1,\ldots,N.
\end{equation}
For $N=0$, we put $\tau_n^t=\sigma_n^t=1$.
Then the main result of this article is given as follows:
\begin{theorem}\label{thm:main}
For $\tau_n^t$ defined above, the functions
\begin{equation}
 A_n^t=-\mu_t^{-1}\frac{\tau_n^t \tau_{n+1}^{t+1}}{\tau_n^{t+1}\tau_{n+1}^t},\quad
 B_n^t=-\mu_t\frac{\tau_{n-1}^{t+1} \tau_{n+1}^t}{\tau_n^t \tau_n^{t+1}},\quad
 \lambda_t=\mu_t+\mu_t^{-1}.
 \label{sol dep var tr}
\end{equation}
satisfy the non-autonomous discrete-time Toda lattice equation (\ref{dTL}).
\end{theorem}
As was pointed out in \cite{MT2004a,MT2004b}, the auxiliary $\tau$ function
$\sigma_n^t$ plays an essential role although it does not appear in the
final result.
\begin{proposition}\label{prop:bilinear}
 $\tau_n^t$ and $\sigma_n^t$ satisfy the following bilinear difference
 equations:
\begin{eqnarray}
&& \tau_n^{t-1}\tau_n^{t+1}-\tau_n^t\sigma_n^t
 =\mu_{t-1}\mu_t(\tau_{n-1}^{t+1}\tau_{n+1}^{t-1}-\tau_n^t\sigma_n^t),
 \label{sol dTL bilin eq 1}\\
&& \mu_t\sigma_n^t\tau_{n+1}^t-\mu_{t-1}\tau_n^t\sigma_{n+1}^t
 =(\mu_t-\mu_{t-1})\tau_n^{t+1}\tau_{n+1}^{t-1}.
 \label{sol dTL bilin eq 2}
\end{eqnarray}
\end{proposition}
Theorem \ref{thm:main} is a direct consequence of Proposition \ref{prop:bilinear}.
Actually, multiplying $1-(\mu_t\mu_{t-1})^{-1}$ to equation (\ref{sol dTL bilin eq 2}), we have
\begin{equation}
 (\mu_t-\mu_{t-1}^{-1})\sigma_n^t\tau_{n+1}^t-(\mu_{t-1}-\mu_t^{-1})\tau_n^t\sigma_{n+1}^t
 =(\lambda_t-\lambda_{t-1})\tau_n^{t+1}\tau_{n+1}^{t-1}.
 \label{sol dTL bilin eq 3}
\end{equation}
Multiplying equation (\ref{sol dTL bilin eq 3}) by $\tau_n^t \tau_{n+1}^t$
and using equation (\ref{sol dTL bilin eq 1}), we have
\begin{eqnarray}
&& (\tau_{n+1}^t)^2 (\mu_t \tau_{n-1}^{t+1}\tau_{n+1}^{t-1}-\mu_{t-1}^{-1}\tau_n^{t-1}\tau_n^{t+1})
 -(\tau_n^t)^2 (\mu_{t-1} \tau_{n-1}^{t+1}\tau_{n+1}^{t-1}-\mu_t^{-1}\tau_n^{t-1}\tau_n^{t+1})\nonumber\\
&=&(\lambda_t-\lambda_{t-1})\tau_n^t\tau_n^{t+1}\tau_{n+1}^{t-1}\tau_{n+1}^t.
 \label{sol dTL quadlin eq}
\end{eqnarray}
Dividing equation (\ref{sol dTL quadlin eq}) by
$\tau_n^t\tau_n^{t+1}\tau_{n+1}^{t-1}\tau_{n+1}^t$, we obtain the first
equation of equation (\ref{dTL}).
The second equation is an identity under the variable transformation~(\ref{sol dep var tr}).
%
%
\begin{remark}
\begin{enumerate}
 \item If we choose the functions $\varphi_i^t(n)$ and $\psi_i^t(n)$ as
       exponential type functions
\begin{eqnarray}
&& \varphi_i^t(n)=\alpha_i p_i^n\prod_{j=t_0}^{t-1}(1-p_i\mu_j)+\beta_i p_i^{-n}\prod_{j=t_0}^{t-1}(1-p_i^{-1}\mu_j),\\
&& \psi_i^t(n)=\alpha_i p_i^n (1-p_i\mu_t)\prod_{j=t_0}^{t-2}(1-p_i\mu_j)
 +\beta_i p_i^{-n} (1-p_i^{-1}\mu_t)\prod_{j=t_0}^{t-2}(1-p_i^{-1}\mu_j),
\end{eqnarray}
respectively, where $\alpha_i$, $\beta_i$, $p_i$ ($i=1,\ldots,N$) are
       parameters, we have the $N$-soliton solution.
As is shown in \cite{Hir2004,OHTI1993}, the $\tau$ fucntions for soliton
       solutions are expressed as Casorati determinants whose entries
       are given by exponential type functions.
 \item In the case where $\mu_t$ is a constant, the bilinear
       equations (\ref{sol dTL bilin eq 1}) and (\ref{sol dTL bilin eq 2}) reduce to
\begin{equation}
 \tau_n^{t-1}\tau_n^{t+1}-(\tau_n^t)^2
 =\mu^2[\tau_{n-1}^{t+1}\tau_{n+1}^{t-1}-(\tau_n^t)^2],
\end{equation}
which is the bilinear equation of the discrete-time Toda
       equation\cite{Hir1977}.
Indeed, the $N$-soliton solution also reduces to that for discrete-time
       Toda equation.
 \item The functions $\varphi_i^t(n)$ ($i=1,\ldots,N$) satisfy the
       spectral problem equation
\begin{eqnarray}
\begin{array}{l}
\medskip
{\displaystyle \varphi_i^{t+1}(n)=-\mu_t\varphi_i^t(n+1)+\varphi_i^t(n),}\\
{\displaystyle (x_i-\lambda_t)\varphi_i^t(n)=-\mu_t^{-1}\varphi_i^{t+1}(n)+\varphi_i^{t+1}(n-1),\quad x_i=p_i+p_i^{-1}.}
\end{array}
\label{Spec eq 2}
\end{eqnarray}
Equation (\ref{Spec eq 2}) is the spectral problem equation (\ref{Spec
       eq 1}) with $A_n^t=-\mu_t^{-1}$, $B_n^t=-\mu_t$ and
       $\lambda_t=\mu_t+\mu_t^{-1}$, which is the simplest solution for
       the non-autonomous discrete-time Toda lattice equation (\ref{dTL}).
\end{enumerate}
\end{remark}
\section{Proof of Proposition \ref{prop:bilinear}}
In this section we prove Proposition \ref{prop:bilinear} by using the
technique developed in \cite{OHTI1993,OKMS1993}.
The bilinear equations (\ref{sol dTL bilin eq 1}) and (\ref{sol dTL
bilin eq 2}) reduce to the Pl\"ucker relations, which are quadratic
identities among the determinants whose columns are properly shifted.
Therefore, we first prepare such difference formulae that express
shifted determinants in terms of $\tau$ or $\sigma$.  For simplicity, we
introduce notation
\begin{equation}
 \tau_n^t =
\left|\begin{array}{ccccc}  0_t & 1_t & \cdots & (N-1)_t \end{array}\right|,
\quad \sigma_n^t
 =\left|\begin{array}{ccccc}  \hat{0}_t & \hat{1}_t & \cdots & \widehat{(N-1)}_t \end{array}\right|,
\end{equation}
where the symbols $k_t$ and $\hat{k}_t$ are column vectors given by
\begin{equation}
 k_t = \left(\begin{array}{c}\varphi_1^t(n+k)\\\varphi_2^t(n+k)\\\vdots\\\varphi_N^t(n+k)\end{array}\right)
,\quad
 \hat{k}_t = \left(\begin{array}{c}\psi_1^t(n+k)\\\psi_2^t(n+k)\\\vdots\\\psi_N^t(n+k)\end{array}\right),
\end{equation}
respectively.  
\begin{lemma}\label{lemma:difference1}
 The following formulae hold:
\begin{eqnarray}
&&\tau_n^t = \left|\begin{array}{ccccc} 0_t & 1_t & \cdots & (N-2)_t & (N-1)_t \end{array}\right|,\label{dif1}\\
&&\tau_n^{t-1} = \left|\begin{array}{ccccc} 0_t & 1_ t& \cdots & (N-2)_t & (N-1)_{t-1} \end{array}\right|,\label{dif2}\\
&&\mu_{t-1}\tau_n^{t-1} = \left|\begin{array}{ccccc} 0_t & 1_t & \cdots & (N-2)_t &(N-2)_{t-1} \end{array}\right|,\label{dif3}\\
&&(\prod_{i=1}^N P_i^t)^{-1}\tau_n^{t+1} 
= \left|\begin{array}{ccccc} \tilde{0}_{t+1} & 1_t & \cdots & (N-2)_t & (N-1)_t\end{array}\right|,\label{dif4}\\
&&(\prod_{i=1}^N P_i^t)^{-1}\mu_t\tau_n^{t+1} 
= \left|\begin{array}{ccccc} \tilde{1}_{t+1}& 1_t & \cdots & (N-2)_t & (N-1)_t\end{array}\right|,\label{dif5}\\
&&(1-\mu_{t-1}\mu_t) (\prod_{i=1}^N P_i^t)^{-1}\sigma_n^{t} 
= \left|\begin{array}{ccccc} \tilde{0}_{t+1}& 1_t & \cdots & (N-2)_t & (N-1)_{t-1}\end{array}\right|,\label{dif6}
\end{eqnarray}
where the symbol $\tilde{k}_t$ is the column vector given by
\begin{equation}
 \tilde{k}_t=\left(
\begin{array}{c}(P_1^t)^{-1}\varphi_1^{t}(n+k) \\(P_2^t)^{-1}\varphi_2^{t}(n+k) \\\vdots\\
(P_N^t)^{-1}\varphi_N^{t}(n) \end{array}\right).
\end{equation}
\end{lemma}
{\it Proof of Lemma \ref{lemma:difference1}:} We first note that 
$\varphi_i^t(n)$ and $\psi_i^t(n)$ also satisfy the linear relations
\begin{eqnarray}
&& \varphi_i^{t+1}(n)=\psi_i^t(n)-\mu_{t-1}\psi_i^t(n+1),
 \label{sol disp rel 4}\\
&& P_i^t \varphi_i^t(n)=\varphi_i^{t+1}(n)-\mu_t \varphi_i^{t+1}(n-1).
 \label{sol disp rel 5}
\end{eqnarray}
which follow from equations (\ref{sol disp rel 1})-(\ref{sol disp rel 3}).
Equation (\ref{dif1}) is nothing but the definition.  Equation (\ref{dif2}) is derived as
follows: Subtracting $(j+1)$-th column multiplied by $\mu_{t-1}$ from
$j$-th column of $\tau_n^{t-1}$ for $j=0,1,\ldots,N-1$, and using
equation (\ref{sol disp rel 1}), we have
\begin{eqnarray*}
 \tau_n^{t-1}&=&\left|
\begin{array}{cccc}0_{t-1} & 1_{t-1} &\cdots & (N-1)_{t-1}\end{array}\right|\\
&=&\left|
\begin{array}{cccc}0_{t-1}-\mu_{t-1}\times 1_{t-1} &  1_{t-1} &\cdots & (N-1)_{t-1}\end{array}\right|\\
&=&\left|
\begin{array}{cccc}0_t &   1_{t-1} &\cdots & (N-1)_{t-1}\end{array}\right|\\
&=&\cdots=\left|
\begin{array}{cccc}0_t &\cdots & (N-2)_{t}& (N-1)_{t-1}\end{array}\right|,
\end{eqnarray*}
which is equation (\ref{dif2}). Moreover, multiplying $\mu_{t-1}$ to the
$N$-th column of right hand side of equation (\ref{dif2}) and using
equation (\ref{sol disp rel 1}), we have
\begin{eqnarray*}
  \mu_{t-1}\tau_n^{t-1}
&=&\left|\begin{array}{cccc}1_t & \cdots & (N-2)_t & \mu_{t-1}\times (N-1)_{t-1}\end{array}\right|\\
&=&\left|\begin{array}{cccc}1_t & \cdots & (N-2)_t & (N-2)_t+\mu_{t-1}\times (N-1)_{t-1}\end{array}\right|\\
&=&\left|\begin{array}{cccc}1_t & \cdots & (N-2)_t & (N-2)_{t-1}\end{array}\right|,
\end{eqnarray*}
which is nothing but equation (\ref{dif3}). Equations (\ref{dif4}) and
(\ref{dif5}) can be proved in similar manner by using equation (\ref{sol
disp rel 5}). Equation (\ref{dif6}) can be proved as follows: first
notice that $\sigma_n^t$ is rewritten as
\begin{equation}
 \sigma_n^t=\left|\begin{array}{cccc} 0_{t+1}&\cdots &(N-1)_{t+1} &\widehat{(N-1)}_t \end{array}\right|,
\label{dif7}
\end{equation}
which is shown in similar manner by using equation (\ref{sol disp rel 4}). We also note
that $\varphi_i^t(n)$ and $\psi_i^t(n)$ satisfy the relation,
\begin{equation}
 (1-\mu_t\mu_{t-1})\psi_i^t(n)=P_i^t\varphi_i^{t-1}(n)+\mu_t\varphi_i^{t+1}(n-1),
\label{sol disp rel 6}
\end{equation}
which can be derived by eliminating $\varphi_i^t(n-1)$ from equation
(\ref{sol disp rel 4}) with $n$ being replaced by $n-1$ and equation
(\ref{sol disp rel 3}). Then multiplying $(1-\mu_t\mu_{t-1})$ to the
$N$-th column of the right hand side of equation (\ref{dif7}) and using
equation (\ref{sol disp rel 6}), we obtain
\begin{eqnarray*}
(1-\mu_t\mu_{t-1})\sigma_n^t
&=&\left|
\begin{array}{cccc} 1_{t+1} & \cdots & (N-2)_{t+1} & (1-\mu_t\mu_{t-1})\times \widehat{(N-1)}_{t-1}\end{array}
\right|\\
&=&
\left|
\begin{array}{cccc}
 \varphi_1^{t+1}(n)&\cdots & \varphi_1^{t+1}(n+N-2) &P_1^t\varphi_1^{t-1}(n+N-1) \\
 \varphi_2^{t+1}(n)&\cdots & \varphi_2^{t+1}(n+N-2) &P_2^t\varphi_2^{t-1}(n+N-1) \\
 \vdots &\cdots& \vdots &\vdots\\
 \varphi_N^{t+1}(n)&\cdots & \varphi_N^{t+1}(n+N-2) &P_N^t\varphi_N^{t-1}(n+N-1) 
\end{array}
\right|\\
&=&\cdots=
\left|
\begin{array}{cccc}
 \varphi_1^{t+1}(n)&P_1^t \varphi_1^{t-1}(n+1)&\cdots  &P_1^t\varphi_1^{t-1}(n+N-1) \\
 \varphi_2^{t+1}(n)&P_2^t \varphi_2^{t-1}(n+1)&\cdots  &P_2^t\varphi_2^{t-1}(n+N-1) \\
 \vdots & \vdots&\cdots &\vdots\\
 \varphi_N^{t+1}(n)&P_N^t \varphi_N^{t-1}(n+1)&\cdots  &P_N^t\varphi_N^{t-1}(n+N-1) 
\end{array}
\right|\\
&=&\prod_{i=1}^N P_i^t~
\left|\begin{array}{cccc}\tilde{0}_{t+1} & 1_{t-1} &\cdots & (N-1)_{t-1}\end{array}\right|,
\end{eqnarray*}
which is equation (\ref{dif6}). This completes the proof of Lemma
\ref{lemma:difference1}. $\square$

Now consider the following identity of $2N\times 2N$ determinant:
\begin{equation}
\left|\begin{array}{c|ccc|ccc|cc}
 \tilde{0}_{t+1} & 0_t & \cdots & (N-2)_t & & \O      &         & (N-1)_t & (N-1)_{t-1}\\
 \hline
 \tilde{0}_{t+1} &     & \O     &         &1_t &\cdots & (N-2)_t & (N-1)_t & (N-1)_{t-1}
 \end{array}\right|=0, \label{det id 1}
\end{equation}
Applying the Laplace expansion to left hand side of equation (\ref{det id 1}) and using
Lemma \ref{lemma:difference1}, we obtain
\begin{eqnarray*}
0&=&\left|\begin{array}{ccccc} \tilde{0}_{t+1}& 0_t& 1_t&\cdots &(N-2)_t \end{array}\right|
\times
 \left|\begin{array}{ccccc} 1_t&\cdots &(N-2)_t &(N-1)_t&(N-1)_{t-1} \end{array}\right|\\
&& + \left|\begin{array}{ccccc} 0_t&1_t&\cdots &(N-2)_t &(N-1)_t \end{array}\right|
\times
 \left|\begin{array}{ccccc} \tilde{0}_{t+1}&1_t&\cdots &(N-2)_t &(N-1)_{t-1} \end{array}\right|\\
&& - \left|\begin{array}{ccccc} 0_t&1_t&\cdots &(N-2)_t &(N-1)_{t-1} \end{array}\right|
\times
 \left|\begin{array}{ccccc} \tilde{0}_{t+1}&1_t&\cdots &(N-2)_t &(N-1)_{t} \end{array}\right|\\
&=&\mu_t(\prod_i^N P_i^t)\tau_{n-1}^{t+1}\times\mu_t\tau_{n+1}^{t-1}
+ \tau_n^t\times (1-\mu_t\mu_{t-1})(\prod_i^N P_i^t)\sigma_n^t
- \tau_n^{t-1}\times (\prod_i^N P_i^t)\tau_n^{t+1},
\end{eqnarray*}
which is the bilinear equation (\ref{sol dTL bilin eq 1}).

The bilinear equation (\ref{sol dTL bilin eq 2}) can be proved by the
similar technique. We prepare the following difference formulae:
\begin{lemma}\label{lemma:difference2}
The following formulae hold.
\begin{eqnarray}
&&\tau_n^t=\left|\begin{array}{ccccc} {0}_{t+1}& 1_{t+1}& \cdots &(N-2)_{t+1}&(N-1)_t \end{array}\right|,
\label{dif8}\\
&& \mu_t\tau_{n+1}^t=
\left|\begin{array}{ccccc} {1}_{t+1}& 2_{t+1}& \cdots &(N-1)_{t+1}&(N-1)_t \end{array}\right|,\label{dif9}\\
&& \sigma_n^t
=\left|\begin{array}{ccccc} {0}_{t+1}& 1_{t+1}& \cdots &(N-2)_{t+1}&\widehat{(N-1)}_t \end{array}\right|,
\label{dif10}\\
&& \mu_{t-1}\sigma_{n+1}^t
=\left|\begin{array}{ccccc} {1}_{t+1}& \cdots &(N-2)_{t+1}& (N-1)_{t+1}&\widehat{(N-1)}_t \end{array}\right|,
\label{dif11}\\
&& (\mu_{t-1}-\mu_t)\tau_{n+1}^{t-1}
=\left|\begin{array}{ccccc} {1}_{t+1}& \cdots &(N-2)_{t+1}& (N-1)_{t}&\widehat{(N-1)}_t \end{array}\right|.
\label{dif12}
\end{eqnarray}
\end{lemma}
\textit{Proof. of Lemma \ref{lemma:difference2}:} Equations (\ref{dif8})
and (\ref{dif9}) are equivalent to equations (\ref{dif2}) and
(\ref{dif3}), respectively. Equation (\ref{dif10}) is the same as
equation (\ref{dif7}). Equation (\ref{dif11}) can be derived by using
equation (\ref{sol disp rel 4}) after multiplying $\mu_{t-1}$ to the
$N$-th column of the right hand side of equation (\ref{dif8}). In order
to prove equation (\ref{dif12}), we note the following relation between
$\varphi_i^t(n)$ and $\psi_i^t(n)$,
\begin{equation}
 (\mu_{t-1}-\mu_t)\varphi_i^{t-1}(n)=\psi_i^t(n-1)-\varphi_i^t(n-1), \label{sol disp rel 7}
\end{equation}
which can be obtained by eliminating $\varphi_i^{t-1}(n)$ from equation
(\ref{sol disp rel 1}) with $t$ being replaced by $t-1$ and equation (\ref{sol disp rel 2}).
Multiplying  $\mu_{t-1}-\mu_t$ on the $N$-th column of
$\tau_{n+1}^{t-1}$ and using equation (\ref{sol disp rel 7}), we obtain
equation (\ref{dif12}). This completes the proof of Lemma
\ref{lemma:difference2}. $\square$

The bilinear equation (\ref{sol dTL bilin eq 2}) is derived 
by applying the Laplace expansion to left hand side of the following identity
\begin{displaymath}
\left|\begin{array}{ccc|ccc|ccc}
 0_{t+1} & \cdots & (N-2)_{t+1} &        & \O     &             & (N-1)_{t+1} & (N-1)_{t} & \widehat{(N-1)}_t\\
 \hline
         &\O      &             &1_{t+1} & \cdots & (N-2)_{t+1} & (N-1)_{t+1} & (N-1)_{t} & \widehat{(N-1)}_t
 \end{array}\right|=0,
\end{displaymath}
and using Lemma \ref{lemma:difference2}. This completes the proof of Proposition
\ref{prop:bilinear} and thus Theorem \ref{thm:main}.
\section{Concluding remarks}\label{Concluding}
%
%
In this article we have presented the $N$-soliton solution for the
non-autonomous discrete-time Toda lattice equation (\ref{dTL}), which can be
regarded as a generalization of the discrete-time Toda equation such that the
lattice interval with respect to time is an arbitrary function in time.

Discrete soliton equations commonly arise as B\"acklund or Darboux type
transformations for contiunous soliton equations. In this
context, a number of iterations of a B\"acklund transformation can be
regarded as the discrete independent variable. The B\"acklund
transformation admits one parameter, playing a role of the lattice
interval, which can be arbitrary function in correspoiding independent
variable. In this sense, discrete soliton equations can be naturally
extended to be non-autonomous (see, for example, \cite{BS1999,Sch2001}). 

As for solutions, however, it seems that only autonomous cases (the
lattice intervals are constants) have been discussed in most of studies.
It was recognized in \cite{KS1991,KOS1993} that the discrete
two-dimensional Toda lattice equation (or equivalently,
the discrete KP
equation) admits a non-autonomous generalization. It is known that various
discrete soliton equations are derived from the discrete KP equation and its
B\"acklund transformations. Therefore it is expected that solutions of
non-autonomous discrete soliton equations are discussed from this point
of view. For example, the solutions of non-autonomous discrete-time
relativistic Toda equation have been constructed in this manner in
\cite{MKO2000}.

However, direct reduction process from the non-autonomous discrete KP
equation might not be sufficient.
As we have shown in this article, in the case of equation (\ref{RTL}),
clever introduction of auxiliary $\tau$ function ($\sigma_n^t$ in this
article) is critical, which does not appear in the autonomous or
continous cases.
Careful investigation of this machinery may lead to various 
generalizations of discrete soliton equations and their solutions.
This problem will be discussed in forthcoming articles.
\section*{Acknowledgement}
The authors would like to thank Dr. Satoshi Tsujimoto and Dr. Ken-ichi
Maruno for valuable discussions. One of the author (K.K) thanks
Prof. Nalini Joshi for hospitality during his stay in University of Sydney.
This work was partly supported by the Grant-in-Aid for JSPS Fellows, The
Ministry of Education, Culture, Sports, Science and Technology, Japan.

\end{document}